\documentclass[%
 reprint,
 amsmath,amssymb,
 prl,
 floatfix, a4paper, aps,
superscriptaddress
]{revtex4-1}

\usepackage{dcolumn}
\usepackage{bm}
\usepackage{siunitx}
\usepackage{comment}
\usepackage{booktabs}
\usepackage{siunitx}
\usepackage{graphicx}
\usepackage{color}
\usepackage{footnote}
\usepackage{array}
\newcolumntype{C}{>{\centering\arraybackslash}p{5.2em}}
\newcolumntype{E}{>{\centering\arraybackslash}p{4.5em}}

\begin{document}


\title{Composition dependence of bulk superconductivity in YFe$_2$Ge$_2$}


\author{Jiasheng Chen}
\affiliation{Cavendish Laboratory, University of Cambridge, Cambridge CB3 0HE, United Kingdom}
\author{Monika B. Gam\.za}
\affiliation{Jeremiah Horrocks Institute for Mathematics, Physics and Astronomy, University of Central Lancashire, Preston, United Kingdom}
\author{Konstantin Semeniuk}
\affiliation{Cavendish Laboratory, University of Cambridge, Cambridge CB3 0HE, United Kingdom}
\author{F. Malte Grosche}
\affiliation{Cavendish Laboratory, University of Cambridge, Cambridge CB3 0HE, United Kingdom}


\date{\today}

\begin{abstract}
\noindent In the layered iron-based superconductor YFe$_2$Ge$_2$, a high Sommerfeld ratio of $\sim \SI{100}{mJ/mol.K^2}$ and a $T^{3/2}$ temperature dependence of the electrical resistivity at low temperature $T$ indicate strong electronic correlations and point towards an unconventional pairing state. We have investigated the role of composition and annealing conditions in optimizing the growth of high-quality YFe$_2$Ge$_2$. Our findings confirm that bulk superconductivity is observed in samples with disorder scattering rates less than $2 k_B T_c/\hbar$. 
Fe deficiency on the Fe site is identified as the dominant source of disorder, which can be minimised by precipitating from a slightly iron-rich melt, following by annealing. 
\end{abstract}


\maketitle

\noindent
The iron-based superconductor YFe$_2$Ge$_2$ \cite{zou14} exhibits strong electronic correlations: 
its Sommerfeld ratio is enhanced by an order of magnitude over density functional theory (DFT) estimates \cite{avila04,subedi14,singh14}, the normal state resistivity $\rho$ follows a non-Fermi liquid temperature dependence, photoemission spectroscopy has revealed renormalised energy bands \cite{xu16}, and it displays enhanced magnetic fluctuations \cite{sirica15,wo18}. Further interest in this material derives from theoretical proposals for the superconducting state, which include $s_\pm$ \cite{subedi14} or triplet pair wavefunctions \cite{singh14} and from its striking similarities to some of the iron pnictide superconductors \cite{chen16,guterding17}.
Moreover, several structurally and electronically related materials have recently been examined, some of which were found to superconduct at low temperatures \cite{felner15,pikul17,chajewski18,samsel18}, including a new iron-based superconductor, LaFeSiH \cite{bernardini17}.

Advanced experiments probing the low temperature state of YFe$_2$Ge$_2$ such as muon spin rotation, penetration depth and quantum oscillation measurements have been held back by the lack of bulk superconducting, high purity single crystals. Although a comprehensive growth study produced flux-grown single crystals \cite{kim15} with comparatively high residual resistivity ratios $\text{RRR}=\rho(\SI{300}{K})/\rho(\SI{2}{K}) \simeq 60$ and sharp resistive superconducting transitions, these did not display a superconducting heat capacity anomaly. Only a second generation of polycrystalline samples with even higher RRR provided thermodynamic evidence for 
bulk superconductivity in YFe$_{2}$Ge$_{2}$ \cite{chen16}.

In order to investigate the origin of this sample dependence, we have conducted a systematic study of the composition phase diagram of YFe$_2$Ge$_2$ and of the effect of annealing conditions on sample quality. Here, we present the key outcomes of this study, which 
may guide the preparation of high quality, bulk superconducting single crystals and  provide new insights into the origin of superconductivity in YFe$_2$Ge$_2$, namely (i) solidification from a slightly iron-rich melt followed by quenching and annealing maximizes Fe occupancy on  the Fe site and produces the samples with the longest electronic mean free path, (ii) $T_c$ is strongly reduced below its optimal value $T_{c0}$ by disorder when the electronic scattering rate approaches $2 k_B T_{c0} /\hbar$, 
(iii) disorder and inhomogeneity induce Griffiths phase signatures, suggesting close proximity of pristine YFe$_2$Ge$_2$ to magnetic instabilities.

\begin{figure}[b]
\includegraphics[width=\columnwidth]{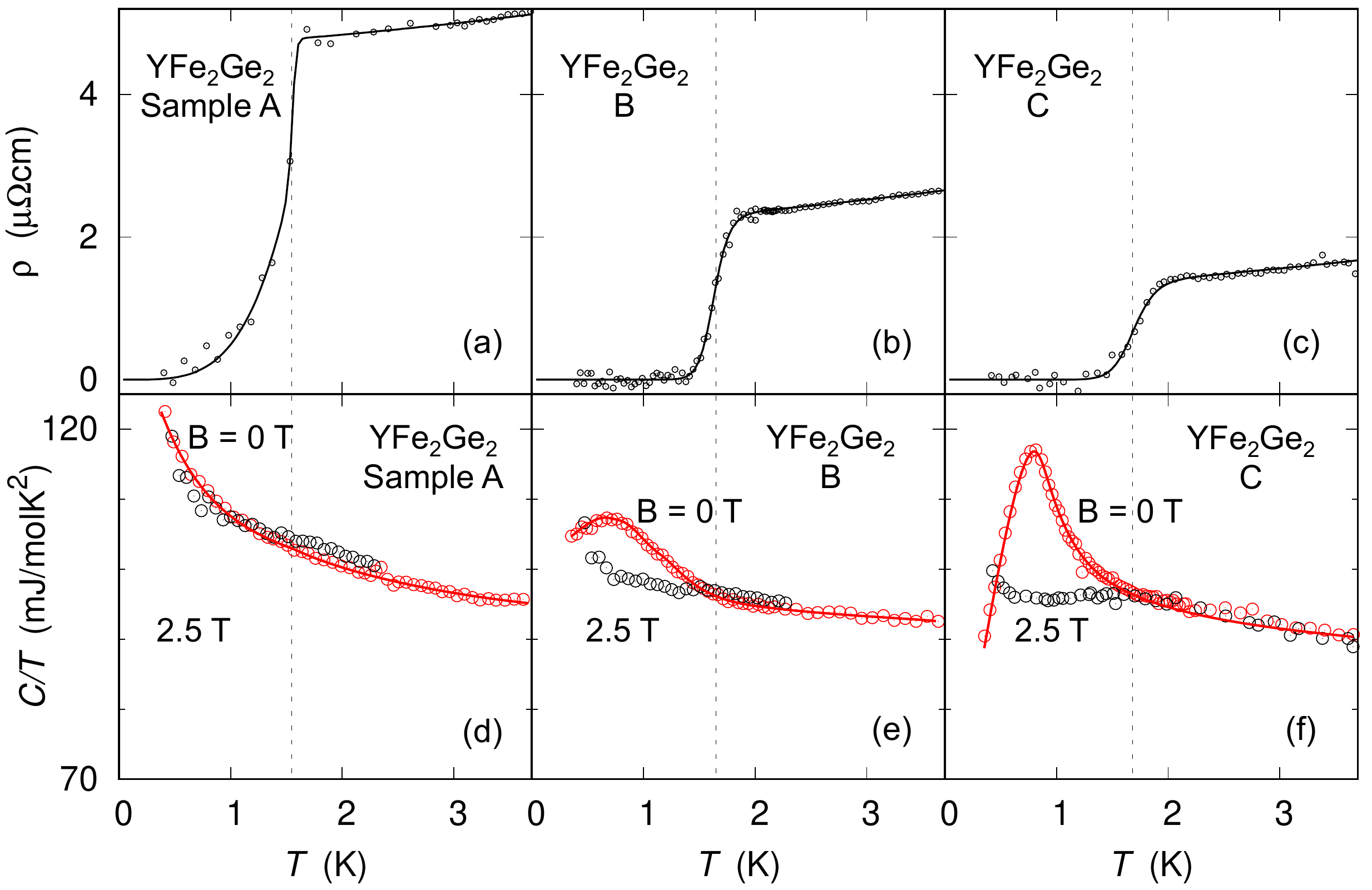}
\caption{\label{FigResC} Electrical resistivity $\rho$ (upper panels a-c) and Sommerfeld coefficient of the heat capacity $C/T$ (lower panels d-f) for three typical samples of YFe$_2$Ge$_2$ with different residual resistivity $\rho_0$. Sample A is as-grown but derives from the same ingot as a bulk superconducting annealed sample, for which data is presented in \cite{chen16}. All three samples show resistive superconducting transitions, but a heat capacity anomaly indicating bulk superconductivity only appears in the purer, annealed samples B and C. When  a superconducting heat capacity anomaly is absent (sample A), $C/T$ displays a slow increase on cooling even in magnetic fields sufficient to suppress $T_c$ fully, suggesting an underlying magnetic contribution. }
\end{figure}


Polycrystalline Y$_{1+x}$(Fe$_{1+y}$Ge$_{1+z}$)$_2$ ingots (with $-0.1 < x, y, z < 0.1$) were grown in a radio-frequency induction furnace on a water-cooled copper boat under a high-purity Ti-gettered Ar atmosphere. To limit the precipitation of stable Y-Ge alloys, Y (3N, Alfa Aesar) and Fe (4N, vacuum remelted, Alfa Aesar) were melted first and thoroughly mixed to form ingots of YFe$_2$. 
Ge (6N, Alfa Aesar), and Y or Fe were then added and melted together with YFe$_2$ to obtain the desired nominal composition. The mass losses due to evaporation were less than $0.3\%$, and homogeneity was was ensured by electromagnetic stirring and repeated flipping and remelting of the ingots. The ingots were quenched to the cooling water temperature within seconds and then heated up again to  near $\SI{1250}{\degreeCelsius}$ for a first annealing step in the induction furnace, which was again followed by rapid quenching. Each of the resulting ingots was mechanically broken up into two halves, one of which was subsequently annealed at $\SI{800}{\degreeCelsius}$ in an evacuated quartz ampoule for 7 days (``annealed''), whereas the other half was investigated without prior annealing (``as-grown''). 
Heat capacity and electrical resistivity were measured with the helium-3 option of the  Quantum Design Physical Properties Measurement System (PPMS) from $300~\text{K}$ to $0.4~\text{K}$ using the pulse-relaxation technique and a standard four-wire ac technique, respectively. The resistivity data were scaled at $\SI{300}{K}$ to the published value of $\SI{190}{\micro \ohm \cm}$ \cite{avila04}.  All the annealed samples and a selection of as-grown samples were measured to check for evidence of bulk superconductivity. 

Powder XRD patterns were collected in the Bragg-Brentano geometry with Cu K$\alpha$ radiation at \SI{40}{kV} and \SI{40}{mA} on a Bruker D8 diffractometer equipped with a Lynxeye XE detector to reduce the effects of Fe fluorescence and K$\beta$ radiation. 
Rietveld refinements of the powder patterns were carried out with FULLPROF. 
Lattice parameters were determined by referring to an internal Ge standard and using the Le Bail method. Multiple measurements were performed on selected batches of representative samples, in order to obtain estimates of the typical variation of lattice parameters. 
Energy dispersive spectroscopy (EDS) was measured with Oxford X-Max detector in an FEI/Philips XL-30 ESEM at \SI{30}{kV} and analyzed with INCA software.


Key features of resistivity and heat capacity data are illustrated in Fig.~\ref{FigResC}.
The as-grown (unannealed) sample A shows a resistive superconducting transition, but no superconducting anomaly in the heat capacity (Fig.~\ref{FigResC} (a, d)). This sample derives from the same ingot as the annealed sample that showed the superconducting heat capacity anomaly in \cite{chen16}. We find more generally that unannealed samples have low RRR values, and while some show a resistive transition all lack superconducting heat capacity anomalies \cite{SuppMat}. 
By contrast, all of the annealed samples show resistive superconducting transitions with varying $T_c$, but not all exhibit signatures of bulk superconductivity in their specific heat (e.g. \cite{zou14}). 
Distinct heat capacity anomalies, namely broad jumps in $C(T)/T$ near $\SI{1}{\kelvin}$ with a peaking at about 20\% above the normal state values, are observed in high-quality samples with RRR above 120 (Fig.~\ref{FigResC}c, f and \cite{chen16}). 
Less prominent anomalies with peaks roughly 10\% above the normal-state $C(T)/T$ can be found in samples with RRR ranging from 60 to 120, 
as illustrated in Fig.~\ref{FigResC}(b, e).

\begin{figure}
\includegraphics[width=\columnwidth]{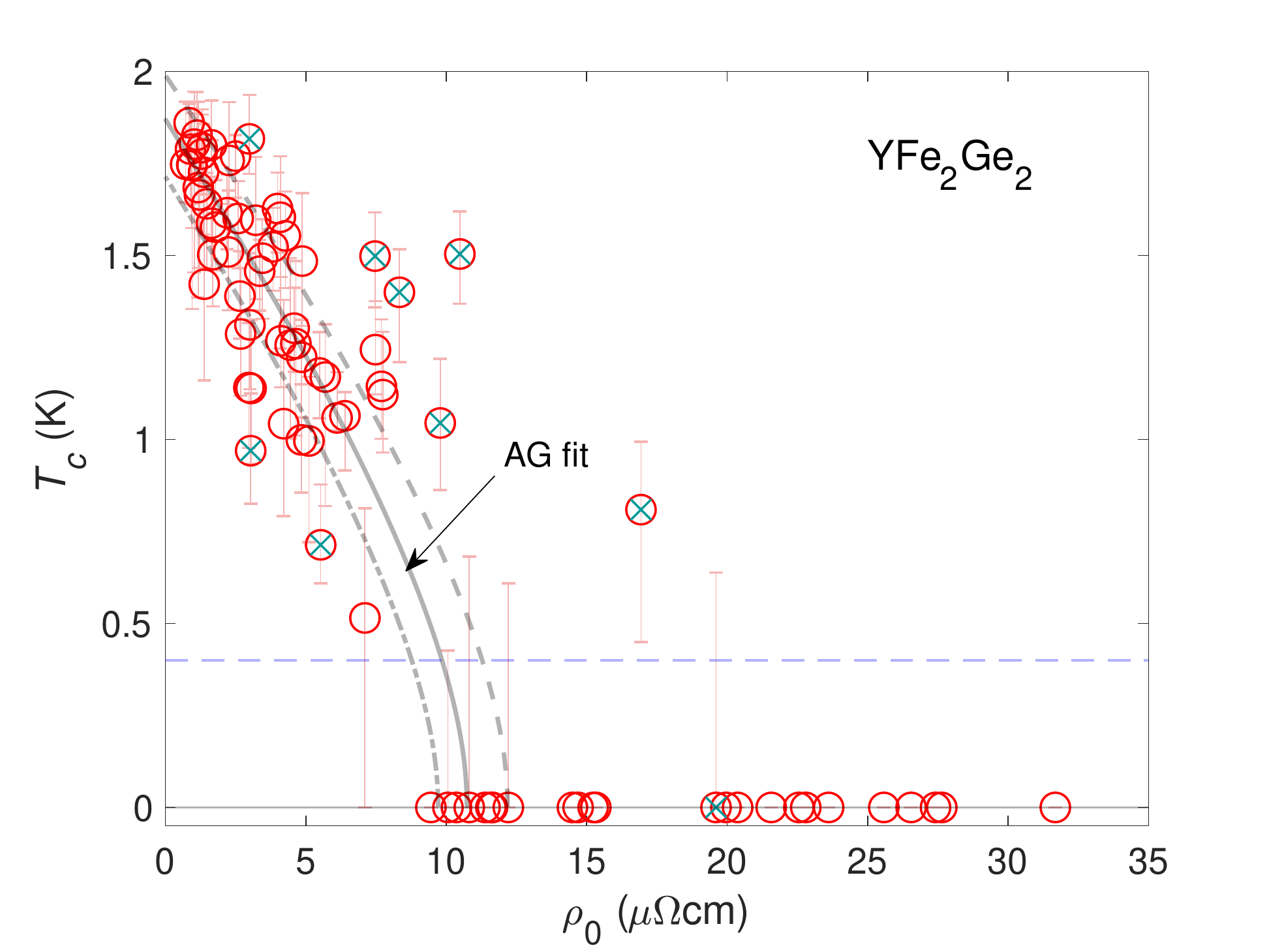}%
\caption{\label{FigTcVsRho0} Resistive transition temperature as a function of the residual resistivity $\rho_0$ for all YFe$_2$Ge$_2$ polycrystalline samples characterized in this study. No data are available below $\SI{0.4}{\kelvin}$ (indicated by the blue dotted line). 
Mid-points of the resistive transitions are shown as circles, with vertical errorbars illustrating the transition widths  (determined by an 80\%/20\% criterion). Solid, dash and dash-dot lines give least-square fits to of the Abrikosov-Gor'kov expression (AG, see text) to the 50\%, 80\% and 20\% $T_c$ points, ignoring those outliers indicated by crosses.}
\end{figure}

The influence of disorder scattering on superconductivity in YFe$_2$Ge$_2$ can be examined quantitatively using the large number of samples  (annealed and as-grown) prepared from more than 20 ingots grown for this study. 
The dependence of the resistive $T_c$ on residual resistivity $\rho_0$ is summarised in Fig.~\ref{FigTcVsRho0}, which illustrates that the data can be modelled by the implicit  Abrikosov-Gor'kov expression \cite{abrikosov60}  
\begin{equation}
\ln\left(\frac{T_{c0}}{T_c}\right) = \Psi\left(\frac{1}{2}+\frac{\alpha T_{c0}}{2\pi T_c }\right) - \Psi\left(\frac{1}{2}\right) ~,
\end{equation}
\noindent where $T_c$ and $T_{c0}$ are the actual transition temperature and the optimal transition temperature without impurity scattering, respectively, $\alpha \propto \rho_0$ measures the pair-breaking effect of impurity scattering and $\Psi(z)$ is the digamma function. 
This approach has been found to describe the experimental data on cuprates 
\cite{radtke93,blackstead95}, the spin-triplet superconductor Sr$_2$RuO$_4$
\cite{mackenzie98,mao99} and the heavy fermion superconductor CeCoIn$_5$ \cite{bauer06}. 
Impurity scattering is expected to suppress an unconventional pairing state, when the pair-breaking parameter $\alpha = \frac {1}{2} \frac{\hbar \tau^{-1}}{k_B T_{c0}}$ approaches $1$ \cite{radtke93, kitaoka94, millis88}, where, $\tau^{-1}$ is the quasiparticle scattering rate. 
Our data suggest an optimal $T_{c0}$ of $\SI{1.87}{\kelvin}$  and show a clear trend for $T_c$ to diminish with increasing $\rho_0$ and superconductivity to be suppressed when $\rho_0 > \SI{10.7}{\micro\ohm\cm}$. 

The scattering rate can be estimated from $\rho_0$ using the Drude result $\tau^{-1}=\epsilon_0 \Omega_p^2 \rho_0 $, where $\Omega_p$ is the renormalised plasma frequency, which is reduced with respect to the bare plasma frequency obtained from a DFT calculation, $\Omega_p^{(0)}$, by the ratio of effective mass $m^*$ over band mass $m_0$: $\Omega_p^2 = (\Omega_p^{0})^2 \frac{m_0}{m^*}$. Estimating $(\Omega^{(0)}_p)^2 = ((\Omega^{(0)}_x)^2 + (\Omega^{(0)}_y)^2 + (\Omega^{(0)}_z)^2)/3 \simeq (\SI{3.43}{eV}/\si{\hbar})^2$ on the basis of DFT calculations \cite{singh14}, and taking the mass enhancement from the ratio of experimental Sommerfeld coefficient $\gamma_{exp}\simeq \SI{100}{\milli\J/\mol\K^2}$ over its DFT counterpart $\gamma_0 \simeq \SI{12.4}{\milli\J/\mol\K^2}$ \cite{singh14} to be $\frac {m^*}{m_0}\simeq 8$,  we find that $\hbar \tau^{-1} =  \rho_0 \epsilon_0 \hbar \Omega_p^2 = \SI{0.197}{\milli\eV} (\rho_0/\si{\micro\ohm\cm})$. For an optimum $T_{c0} \simeq \SI{1.87}{K}$, this gives $\alpha= 0.62 (\rho_0/\si{\micro\ohm\cm})$, which would suggest that superconductivity should already be fully suppressed when $\rho_0$ exceeds about $\SI{1.6}{\micro\ohm\cm}$. This contrasts with the threshold of $\SI{10}{\micro\ohm\cm}$ for full resistive transitions. The resistive transition, although eventually suppressed, is therefore more robust than might be expected, which may indicate that percolating superconducting paths through high purity regions of a sample can be found even in samples in which the averaged resistivity ratio is comparatively low. The experimental observation that residual resistivities of less than $\SI{2}{\micro\ohm\cm}$ are required to observe {\em thermodynamic} signatures of the superconducting phase transition, by contrast, is fully in line with this analysis.



\begin{figure}
\includegraphics[width=\columnwidth]{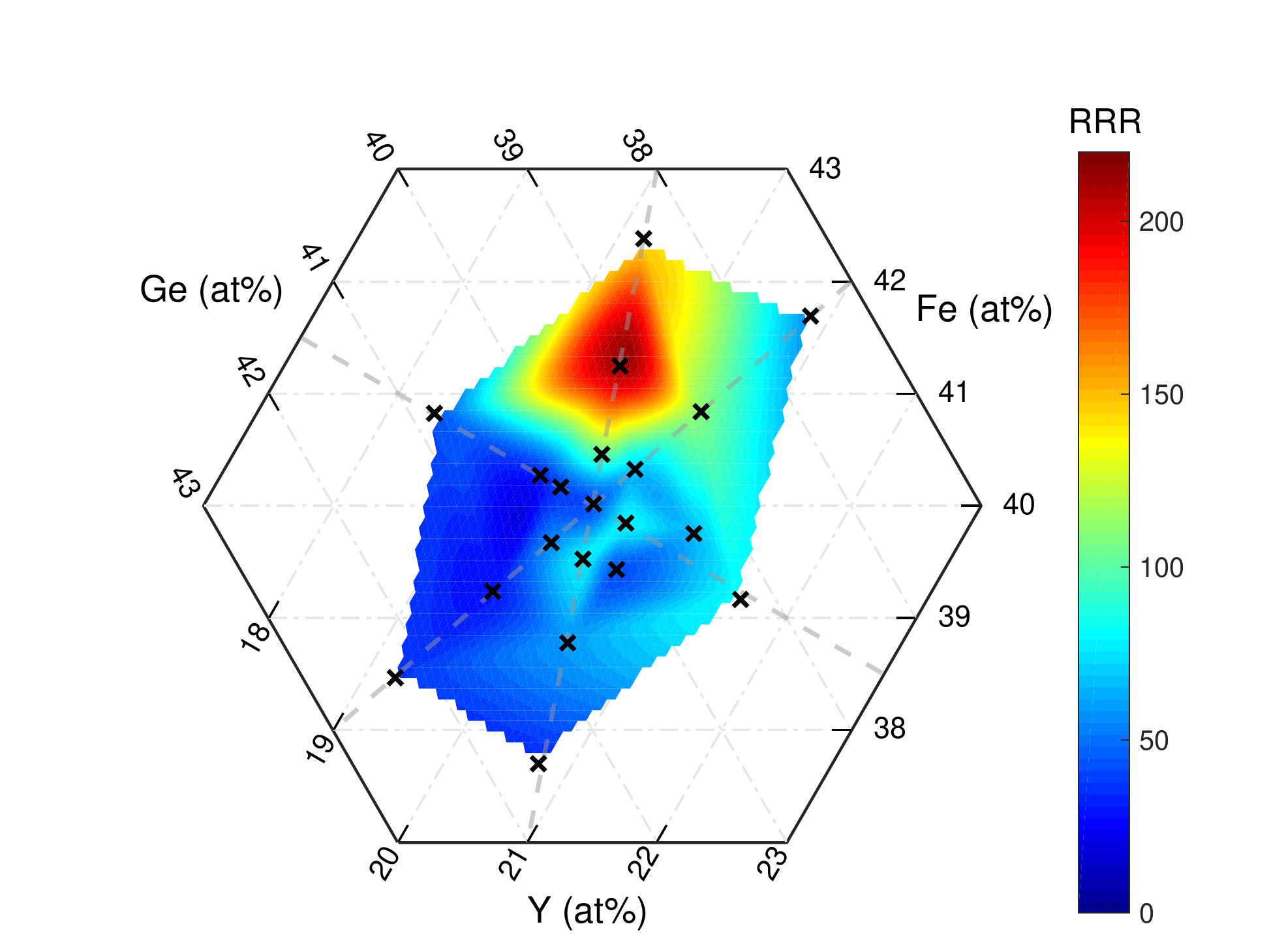}
\includegraphics[width=\columnwidth]{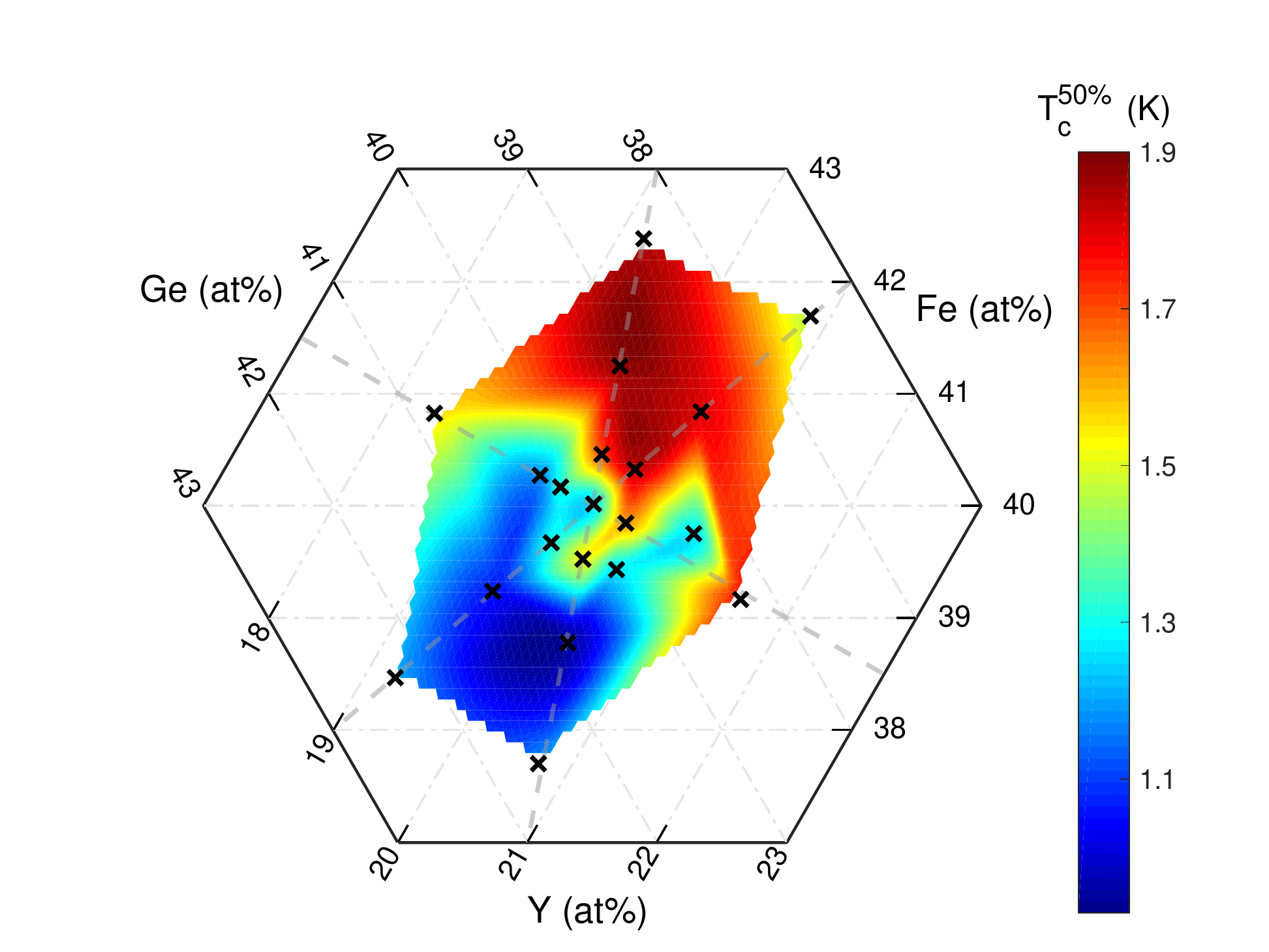}

\caption{\label{FigRRRTc} Residual resistance ratio RRR $=\rho (\SI{300}{\kelvin}) / \rho (\SI{2}{K})$ (upper panel) and mid-point of the resistive superconducting transition (lower panel) of annealed, polycrystalline Y$_{1+x}$(Fe$_{1+y}$Ge$_{1+z}$)$_2$ on a ternary diagram.  Data shown in the ternary plots represent the highest values of RRR or $T_c$ observed for nominal compositions marked by crosses, and interpolated in between (see \cite{SuppMat} for tables of compositions and for data on both as-grown and annealed samples).
}
\end{figure}


Having established the central role of impurity scattering in suppressing superconductivity in YFe$_2$Ge$_2$, we now examine the influence of growth conditions on sample quality. 
The variation of resistance ratio RRR and resistive $T_c$ with nominal composition is summarised in Fig.~\ref{FigRRRTc}, with additional detail, in particular the corresponding data for as-grown samples available in \cite{SuppMat}. 
Our data indicate that the primary influence on sample quality is the ratio of Fe vs. Ge content in the melt: along the line of constant Y content (running diagonally to the top-right of the figure), both RRR and $T_c$ show the largest variation, and growth from an Fe-rich, Ge-poor melt results in higher RRRs and $T_c$s. In particular, the highest RRR of 211 was observed in a sample selected from the annealed Y(Fe$_{1.05}$Ge)$_2$ ingot (\# 34) which also exhibits an enhanced $T_c$ of $\SI{1.87}{\kelvin}$ \cite{chen16}.

\begin{figure}
\includegraphics[width=\columnwidth]{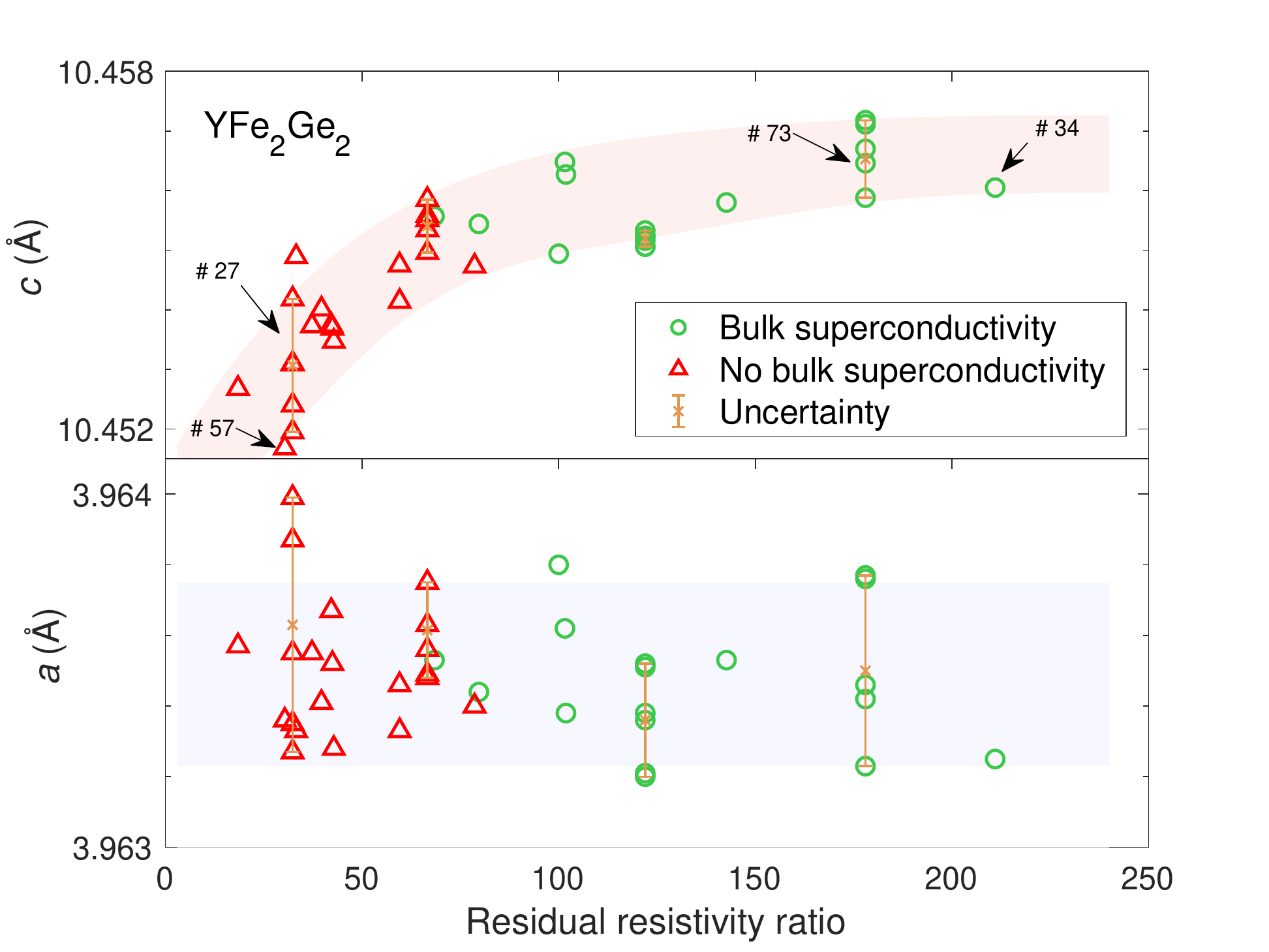}%

\caption{\label{c_and_a_vs_RRR} Lattice constants $c$ (upper panel) and $a$ (lower panel) of the majority phase in each batch of sample as obtained by XRD refinement, versus the corresponding RRR values. Green circles (red triangles) indicate that superconducting heat capacity anomalies have (have not) been observed. Errorbars are estimated from repeated measurements on the selected batches.}
\end{figure}

The nature of the YFe$_2$Ge$_2$ samples grown from melts of varying composition has been analysed by powder x-ray diffraction and energy dispersive spectroscopy (EDS). Because YFe$_2$Ge$_2$ has a narrow homogeneity range, as documented by the high RRR values achieved after annealing, alien phases must be present in nominally off-stoichiometric samples, and have indeed been detected and characterised \cite{SuppMat}. However, the observed signatures of superconductivity in YFe$_2$Ge$_2$ cannot be attributed to these alien phases, because volume superconductivity has also been detected in samples with negligible alien phase content and in samples in which the alien phases are known to be non-superconducting (e.g. \cite{chen16}).
X-ray lattice constant measurements reveal a clear correlation between the lattice parameters of samples with varying nominal composition and their corresponding maximal RRR (Fig.~\ref{c_and_a_vs_RRR}). 
Whereas the $a$-axis lattice parameter is the same in different ingots, a trend towards larger $c$-axis lattice parameters is observed for higher-quality samples, which also tend to show bulk superconductivity.


In order to investigate a possible link between $c$-axis lattice parameter and sample composition, we performed EDS measurements on samples showing the smallest (\# 27 and 57) and the largest (\# 34 and 73) $c$-parameters. As shown in Tab.~\ref{EDSshort}, the Fe:Ge ratios are noticeably higher in \# 34 and 73, both of which have higher RRR and larger $c$-parameters, than in \# 27 and 57. Samples with RRR as high as 200 must be close to the ideal stoichiometry. The EDS results therefore point towards a lack of Fe in samples with lower RRR, which could arise from substitution of Fe by Ge on Fe sites or from Fe vacancies. Because the metallic radius of Fe is slightly larger than the covalent radius of Ge, either possibility would explain the reduction of the $c$-axis lattice parameter in the lower quality, Fe-poor samples. Based on these findings we expect that the homogeneity range of YFe$_2$Ge$_2$ is elongated along the Fe/Ge (constant Y content) axis, but its Fe-rich border passes very close to the stoichiometric position, whereas its Ge-rich border is well away from stoichiometry. The highest quality samples are then grown from an iron-rich melt. 

\begin{table}
\begin{ruledtabular}
\begin{tabular}{l*{4}{C}}

  Ingot No. & \#  27  & \#  57 & \#  34 & \#  73 \\
  Fe:Ge & 0.965(2) & 0.967(2) & 0.984(2)& 0.985(2) \\
  
\end{tabular}
\end{ruledtabular}
\caption{Fe/Ge concentration ratios determined by EDS \cite{SuppMat} for four YFe$_2$Ge$_2$ ingots of varying RRR, as shown in Fig.~\ref{c_and_a_vs_RRR}.}
\label{EDSshort}
\end{table}





The observation that the highest $T_c$ and the clearest thermodynamic signatures of superconductivity are found in the samples with the highest RRR points towards  unconventional superconductivity. This, and the unusually high Sommerfeld coefficient in YFe$_2$Ge$_2$ prompt the question whether this material is close to a magnetic quantum critical point (QCP) and whether magnetic order could be induced in it. Varying the Fe/Ge ratio opens up the possibility of tuning the electronic and magnetic properties of YFe$_2$Ge$_2$, which according to DFT calculations \cite{subedi14,singh14,wang16} and recent NMR and neutron scattering experiments \cite{srpcic17,wo18} is finely balanced close to several types of magnetic order. In contrast to CeCu$_2$Si$_2$, which can be tuned between magnetically ordered and fully superconducting low temperature states by varying the sample composition \cite{gegenwart98}, no magnetic transitions have been observed in any of our samples of YFe$_2$Ge$_2$. This is consistent with the comparatively high RRR observed in all annealed samples, which points towards a homogeneity range that is too narrow to include the magnetic sector of the low temperature phase diagram. A magnetic quantum critical point can be accessed by doping, as in the alloying series (Lu/Y)Fe$_2$Ge$_2$ \cite{ran11}, but no doped samples have shown any signatures of superconductivity, consistent with the view that disorder scattering rapidly suppresses superconductivity in YFe$_2$Ge$_2$. 

Although there is no evidence for long-range magnetic order in any of our samples, we observe pronounced upturns in $C(T)/T$ below $\SI{2}{\kelvin}$ (Fig.~\ref{FigResC} (d)) in the more disordered samples ($\text{RRR}< 60$). Similar upturns in $C/T$ are also evident in earlier samples of YFe$_2$Ge$_2$ which do not display bulk superconductivity \cite{zou14,kim15,avila04}, and they are found in as-grown samples with iron-rich as well as iron-poor nominal composition, indicating that this phenomenon is not primarily a consequence of composition-tuning. Because these upturns persist under applied magnetic field up to $\SI{2.5}{\tesla}$, above the resistive upper critical field of YFe$_2$Ge$_2$, they are unlikely to be caused by a superconducting transition. 
The low-$T$ heat capacity upturns are absent in the annealed samples with the lowest disorder levels, which also show bulk superconductivity (e.g. Figs.~\ref{FigResC} (e, f)), even if in some cases upturns were observed in the more disordered, as-grown parts of ingots with the same composition. The $C(T)$ upturns therefore cannot be attributed to a clean-limit magnetic QCP but may instead be caused by magnetically ordered rare regions, or Griffiths-phase phenomena, which arise naturally in disordered and inhomogeneous samples near a clean-limit QCP \cite{vojta06}. Although it has not been possible to access a magnetically ordered state by shifting the Fe/Ge ratio within the very narrow homogeneity range of YFe$_2$Ge$_2$, these  findings on disordered samples are consistent with the view that this material is very close to a magnetic instability.

Our study demonstrates that the lowest disorder polycrystals of YFe$_2$Ge$_2$ can be grown by shifting the Fe/Ge ratio in the melt to favour full Fe occupancy on the Fe sites. Followed by annealing, which minimises anti-site disorder, this produces samples with residual resistivities as low as $\sim \SI{1}{\micro\ohm\cm}$, which exhibit superconducting anomalies in the heat capacity as well as in the resistivity. The strong correlation between the residual residual resistivity and $T_c$ (Fig.~\ref{FigTcVsRho0}) found by studying dozens of samples with a wide range of nominal compositions, both as-grown and annealed, is reminiscent of well-known unconventional superconductors such as Sr$_2$RuO$_4$ \cite{mackenzie98},  YBa$_2$Cu$_3$O$_{7-\delta}$ \cite{rullier03} and CeCoIn$_5$ \cite{bauer06}. Even the purest samples display a striking separation between the resistive $T_c$, which is as high as $\simeq \SI{1.87}{K}$, and the heat capacity anomaly, which occurs below about $\SI{1.1}{K}$. This separation may be attributed to spatial inhomogeneity within the sample, but the nature of this inhomogeneity -- whether composition, disorder level or strain (e.g. \cite{bachmann18}) -- requires further investigation.

\begin{acknowledgments}
We thank, in particular, C. Geibel, F. Steglich, G. Lonzarich, S. Dutton and J. Baglo for helpful discussions. The work was supported by the EPSRC of the UK (grants no. EP/K012894 and EP/P023290/1) and by Trinity College. 

\end{acknowledgments}

\bibliography{YFGRefs}

 \clearpage
 \onecolumngrid
\pagenumbering{arabic}
\renewcommand\thefigure{S\arabic{figure}} 
\setcounter{figure}{0} 
\renewcommand\thetable{S\arabic{table}} 
\setcounter{table}{0}

\begin{center}
{\bf{Supplementary Material for \\ ``Composition dependence of bulk
    superconductivity in YFe$_2$Ge$_2$''}}\\
\vspace{4em}
\end{center}
\twocolumngrid

\noindent
This material tabulates the nominal compositions of all the
samples for which data is shown in the main manuscript and provides
impurity phase contents for these samples as extracted from powder
x-ray diffraction. Based on this data, it presents a tentative metallurgical
phase diagram for the Y-Fe-Ge system. It also tabulates the energy dispersive
spectroscopy (EDS) data for four of the ingots, for which extracts of
this data is presented in the main paper, and it includes low
temperature resistivity data for four annealed and four as-grown
samples, showing the evolution of the resistive superconducting
transition with disorder level.


\section{Ternary phase diagram}
\noindent
In order to sample the ternary phase diagram near the stoichiometric
composition of YFe$_2$G$_2$, we varied the nominal stoichiometry by approximately $\pm 2\%,~ \pm 5\%$ and $\pm 10\%$ in each of the elements, as well as preparing some nominally stoichiometric ingots. 
 Table \ref{XRD} lists the ingot numbers, their nominal compositions and the impurity phase contents estimated from powder x-ray diffraction. The formation of impurity phases from off-stoichiometric melts is a natural consequence of the narrow homogeneity range of YFe$_2$Ge$_2$. 

\begin{figure}
\includegraphics[width=\columnwidth]{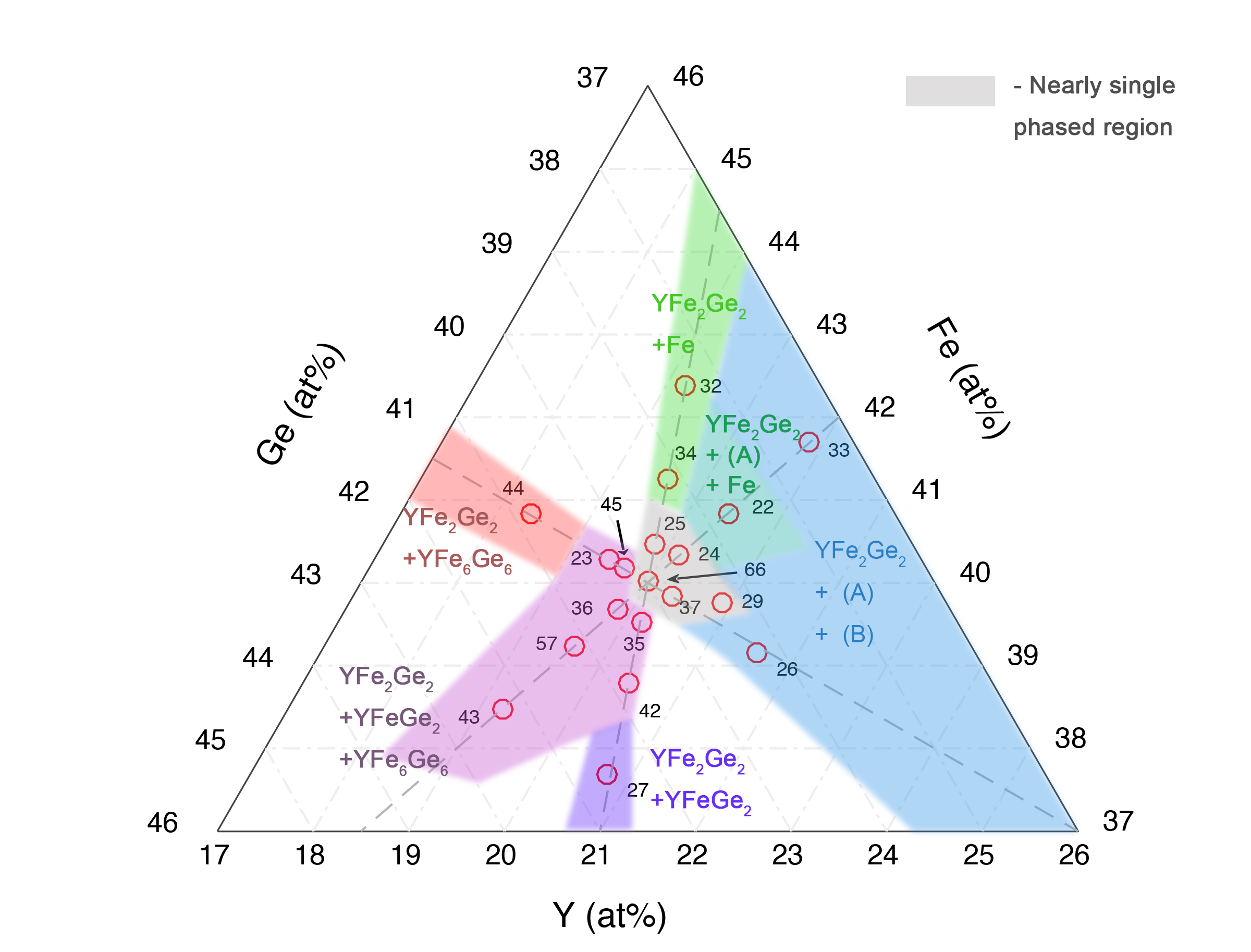}%
\caption{\label{FigPhaseDig} Proposed metallurgical phase diagram of Y-Fe-Ge near stoichiometric YFe$_2$Ge$_2$ inferred from powder X-Ray diffraction results in Tab.~ \ref{XRD}.}
\end{figure}

Figure \ref{FigPhaseDig} summarises this data graphically, showing the sectors in the nominal composition ternary phase diagram, in which certain alien phases were detected. 
Away from the 1-2-2 composition, five main secondary phases are found
in different regions of the phase diagram: YFe$_6$Ge$_6$,
YFe$_{1-x}$Ge$_2$, an Fe-rich bcc ferromagnetic Fe/Ge alloy (just
denoted 'Fe' in what follows) and two
phases (A) and (B) whose exact structures are not yet
identified. Phase (A) shows spectral peaks close to those of
YFe$_4$Ge$_2$ \cite{schobinger01} but with mismatching intensities,
whereas phase (B) is likely to be a compound with higher yttrium
concentration than YFe$_2$Ge$_2$ according to our EDS studies. No
record of phase (B) can be found in the literature. As described in
the main text, the highest quality samples -- those that display the
highest residual resistivity ratio RRR and the highest superconducting
transition temperatures $T_c$ -- form from nominally Fe-rich
compositions. 
They are mainly located in the nearly single-phased region and the
region for which Fe is the impurity phase.

\begin{figure}[b]
\includegraphics[width=\columnwidth]{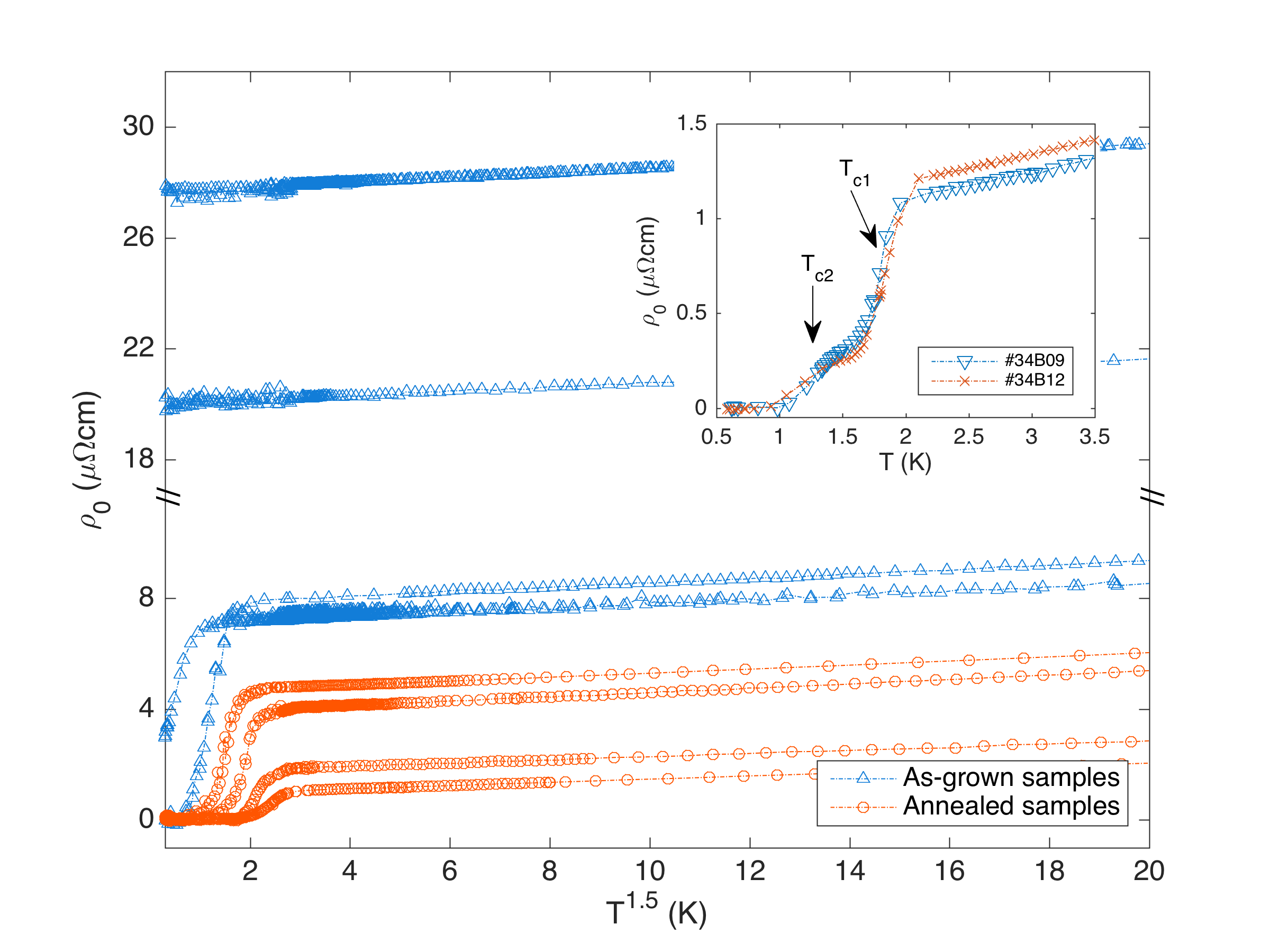}
\caption{\label{FigRes}Electrical resistivities for a representative set of YFe$_2$Ge$_2$ polycrystalline samples versus temperature, variations in $T_c$ and $\rho_0$ and a T$^{3/2}$ normal state temperature dependence. (inset) Electrical resistivities of two samples that showed two-step superconducting transitions.}
\end{figure}

\begin{figure}[b]
\includegraphics[width=\columnwidth]{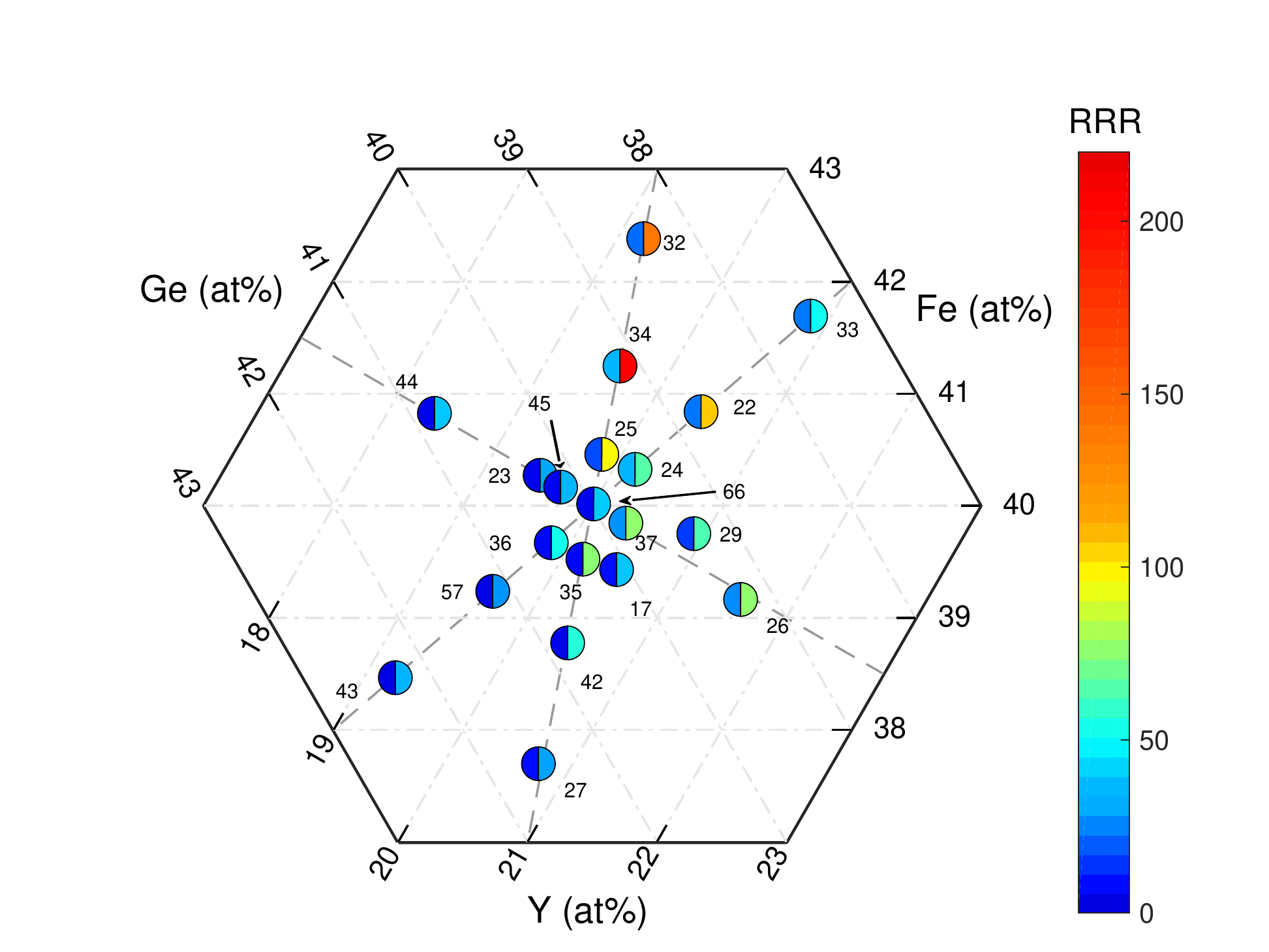}
\includegraphics[width=\columnwidth]{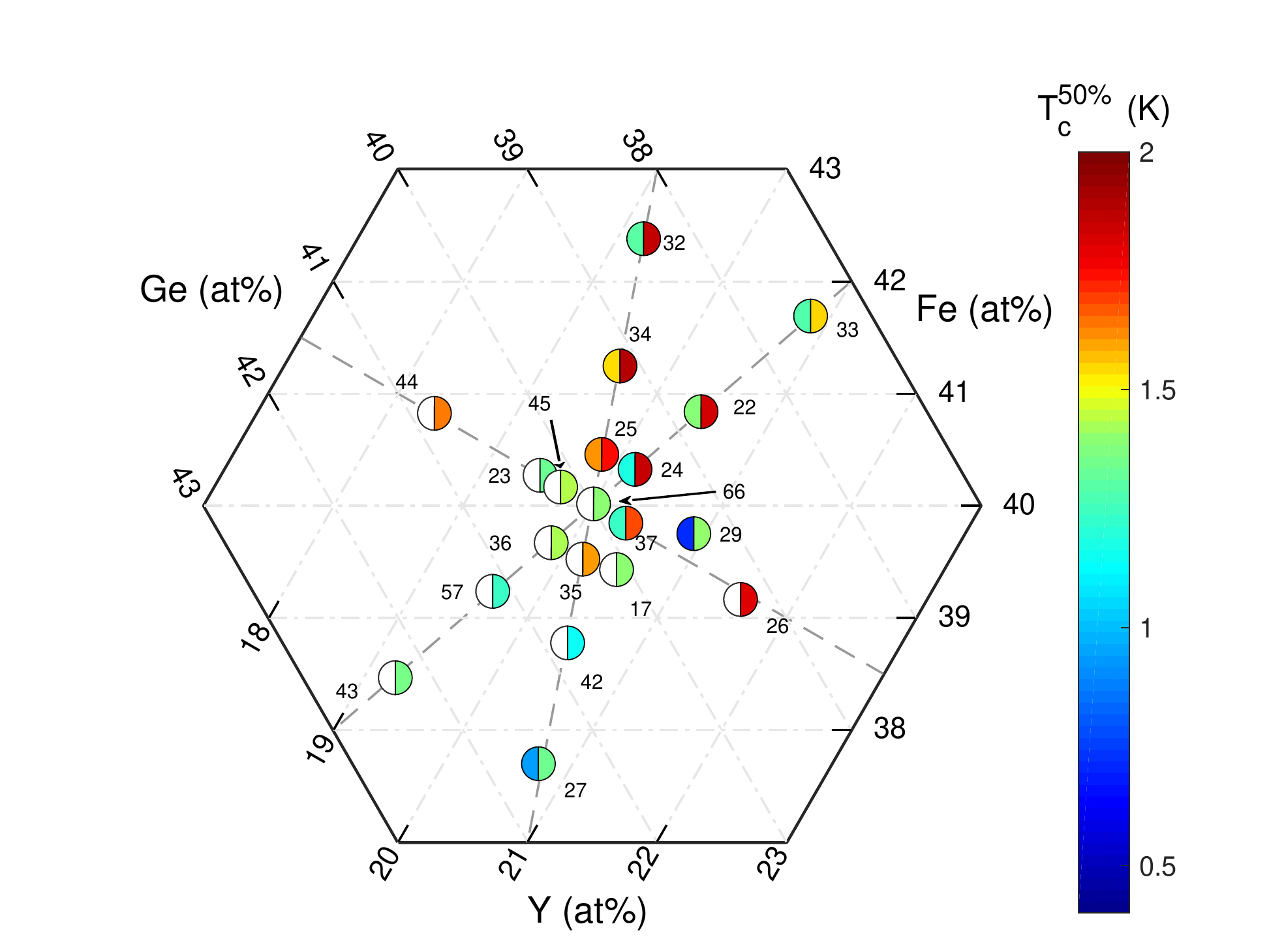}
\caption{\label{FigTernary}Ternary phase diagrams of sample RRR (upper panel) and mid-point resistive $T_c$ (lower panel) for both as-grown (left half of circle) and annealed samples (right half of circle). White color refers to absence of a resistive superconducting anomaly, and the numerical labels denote the ingot number, as used for instance in the tables of compositions and EDS data, Tab.~\ref{XRD} and \ref{EDS}.}
\end{figure}

\section{Comparison of electrical resistivity data from annealed and as-grown samples}
\noindent
The residual resistance ratio represents a rough measure of the differences in lattice disorder of the main phase due to anti-site, interstitial or vacancy defects. As we can see from Fig.~1 in the main paper and data presented in \cite{chen16}, despite having the highest range of RRR recorded so-far in YFe$_2$Ge$_2$ and the sharp resistive transitions near $\SI{1.8}{\kelvin}$, the bulk transitions of these samples appear to be rather broad. This broadness can be attributed to inhomogeneity of disorder and impurity within each batch of sample which, as indicated by Fig.~2 in the main manuscript, can result in a distribution of $T_c$ in each sample. In the inset of Fig.~\ref{FigRes}, $\rho(T)$ of two samples from the highest-quality batch (\# 34) are plotted, both showing two distinct transitions. These two-step transitions are only observed in samples with dimensions of the order $\SI{200}{\micro\meter} \times \SI{100}{\micro\meter} \times \SI{50}{\micro\meter}$. In larger samples, by contrast, we sometimes find relatively broad resistive transitions as illustrated by the wide errorbars in Fig.~2 of the main manuscript.

 The striking effect of heat treatment at $\SI{800}{\degreeCelsius}$ is illustrated by both the appreciable increases in RRRs and $T_c$s after annealing and the appearance of superconductivity in samples that did not superconduct (down to $\SI{0.4}{K}$) prior to the annealing. 
The contrast between annealed and as-grown samples is also highlighted in Fig.~\ref{FigTernary}, which summarises the RRR and mid-point resistive $T_c$ from all samples. These improvements in sample qualitities can be understood in terms of the releasing of disorder created due to fast quenching during growth, such as the effects of dislocations, strains and anti-site defects. 
Further annealing at $\SI{800}{\degreeCelsius}$ for another 7 days did not result in noticeable changes in sample qualities.

\section{Scattering rate and mean free path}
\noindent
The calculation of the scattering rate given in the main text is based on the Drude result $\tau^{-1}  = \epsilon_0 \Omega_p^2 \rho_0$, where the quasiparticle plasma frequency $\Omega_p$ has been corrected with respect to the DFT result to include the mass renormalisation of order 10 in this material. The Abrikosov-Gorkov form for the dependence of $T_c$ on $\tau^{-1}$ given in the main text requires the pair breaking parameter $\alpha = \frac{1}{2}\frac{\hbar \tau^{-1}}{k_B T_{c0}}$ to be less than 1 for superconductivity to occur.

For the calculation in the main text, we have used a published DFT estimate for the plasma frequency, which uses the relation
\begin{equation}
\Omega_{x}^2 = \frac{e^2}{4\pi \epsilon_0 }\frac{1}{\pi^2 m^2} \sum_{n,
  \bf{k}} p_{x; n, {\bf k}} ^2 \delta(\epsilon_{n, {\bf k}} -
\epsilon_F) \; ,
\end{equation}
and likewise for the other principal axes $y$ and $z$. Here, ${\bf
  p}_{n,{\bf k}}$ is the momentum expectation value for states in
band $n$ with crystal momentum $\bf {k}$. The overall unrenormalised plasma frequency
is then estimated by averaging the squared frequencies: $(\Omega_p^{(0)})^2 =
\frac{1}{3} (\Omega_x^2 + \Omega_y^2 + \Omega_z^2)$, and the renormalised plasma frequency is given by $\Omega_p^2 = (\Omega_p^{(0)})^2 \frac{\gamma_{DFT}}{\gamma_{exp}}$, where $\gamma_{DFT}$ and $\gamma_{exp}$ are the DFT-derived and the measured low temperature Sommerfeld coefficients, respectively. Using this criterion, we find that bulk superconductivity should be suppressed when the residual resistivity exceeds $1.6 ~\mu\Omega\text{cm}$.

An independent criterion for the limiting disorder scattering can be obtained by considering the mean free path rather than the scattering rate, and comparing it to the superconducting coherence length: the BCS coherence length $\xi_{BCS} =\frac{\hbar v_F}{\pi \Delta}$, where $v_F$ is the quasiparticle (renormalised) Fermi velocity and $\Delta = \eta k_B T_{c0}$ (with $\eta = 1.76$ in weak-coupling BCS theory) is an estimate of the energy gap, can be compared to the mean free path $\ell = v_F \tau$ to give:
\begin{equation} 
\xi/\ell = \alpha \frac{2 k_B T_{c0}}{\pi \Delta} \simeq \frac{2}{\pi\eta} \alpha 
\end{equation}
Taking the critical value for the pair-breaking parameter $\alpha$ to be $\sim 1$ and $\eta \simeq 2$ then implies that $\ell$ has to be about four times larger than $\xi$ for superconductivity to be observed. As the experimental value for the coherence length based on the observed upper critical field for bulk superconductivity \cite{chen16} are $\xi \simeq \SI{180}{\angstrom}$, bulk superconductivity would then require $\ell > \SI{600}{\angstrom}$. An estimate can be obtained for $v_F$ by combining the density of states per unit volume (from the experimental $C/T$) and the renormalised plasma frequency, via $\left\langle v_F^2 \right\rangle = \frac{3\epsilon_0}{e^2}\frac{\Omega_p^2}{g(E_F)}$ (e.g. \cite{chakraborty76}) as $v_F \simeq \SI{3.3e4}{m/s}$. With the expression for $\tau$ given above, we obtain the estimate
\begin{equation}
\ell = \left(\frac{3}{\epsilon_0 e^2 g(E_F) \Omega_p^2}\right)^{1/2} \frac{1}{\rho_0} = \SI{1100}{\angstrom} (\rho_0/\si{\micro\ohm\cm})^{-1} ~,
\end{equation}
comparable to the value of $\SI{1500}{\angstrom} (\rho_0/\si{\micro\ohm\cm})^{-1}$ given in \cite{chen16}. For a required mean free path of $\SI{500}{\angstrom}$, this translates to a critical resistivity $\rho_0 = \SI{1.8}{\micro\ohm\cm}$, slightly larger than the value found in the main text by comparing the relaxation rate to $k_B T_{c0}$ directly.




\section{Energy dispersive spectroscopy (EDS)}
\noindent EDS measurements were performed on a subset of ingots, namely two with very high RRR (\# 73 and 34) and two with low RRR (\#27 and 57). On each sample, spectra were obtained at around 80 spots of size $\sim\SI{400}{\micro\meter^2}$, free of secondary phases, distributed over different surface sites of each polished sample. Measurements were calibrated against elemental standards, giving an absolute uncertainty in the composition of the order of 2 at.\%, as illustrated by EDS measurements on polycrystalline FeGe$_2$ and Y$_5$Ge$_3$ samples (Tab.~\ref{EDS}). Relative {\em changes} in composition can be measured to a much higher precision, however. All EDS measurements were done under the same conditions, and for the analysis of the Fe/Ge ratio, the K-lines of Fe and Ge were used, which are close in energy. Using the EDS measurements on FeGe$_2$ as a standard in order to recalibrate the EDS Fe/Ge ratios produces the values displayed in Tab.~1 in the main manuscript.

\onecolumngrid

\begin{table*}
\begin{ruledtabular}
\begin{tabular}{c*{8}{E}}
    Ingot No. \#&\multicolumn{3}{>{\centering\arraybackslash}m{.3\linewidth}}{Nominal Composition  Y$_{1+x}$[Fe$_{1+y}$Ge$_{1+z}$]$_2$}
  &\multicolumn{5}{>{\centering\arraybackslash}m{.5\linewidth}}{Impurity Phases (wt.\%)}\\
  \cmidrule(r){2-4} \cmidrule(r){5-9}
    &x (\%)&y (\%)&z (\%)&YFeGe$_2$&YFe$_6$Ge$_6$&Fe (bcc)&Phase (A)&Phase (B)\\
    \midrule
  26&10.1  &0     &0.3   &-        &-            &-       &10 - 20&10 - 20    \\
  29&5.2   &0     &-1    &-        & $<1$        &-       & -     & -         \\
  37 {\bf (B)} &2.1   &0     &-0.1  &-        &-            &-       &-      &-          \\
  45&-2.1  &0     &0     &-        &17.5         &-       &-      &-          \\
  23&-3.4  &0     &0     &8.1      &21.7         & -      &-      &-          \\
  44&-10   &0     &0     &-        &32.4         &-       &-      &-          \\
    \cmidrule(lr){1-9}
  32&0     &10.3  &0     &-        &-            &2.7     &-      &-          \\
  34 {\bf (A)} &0     &5.3   &0     &-        &-            &1.5     &-      &-          \\
  25&0     &1.9   &0     &-        &-            &$<1$    &-      &-          \\
  35&0     &-2    &0     &2.7      &4.3          &-       &-      &-          \\
  42&0     &-5.1  &-0.1  &8.1      &4.6          &-       &-      &-          \\
  27&0     &-9.1  &0.2   &13.1     &-            &-       &-      &-          \\
    \cmidrule(lr){1-9}
  43&0     &-0.1  &9     &10.8     &19           &-       &-      &-          \\
  57&0     &0     &4.9   &13.1     &44           &-       &-      &-          \\
  36&0     &-0.1  &2     &1.8      &7.3          &-       &       &-          \\
  24&0     &0     &-2.1  &-        &-            &$<1$    &$<1$   &$<1$       \\
  22&0     &0     &-5.2  &-        &-            &1.1     &$<1$   &-          \\
  33&0     &0     &-10.1 &-        &-            &-       &10 - 20&10 - 20    \\
    \cmidrule(lr){1-9}
  66&0     &0     &0     &-        &6.1          &-       &-      &-          \\
  72\footnotemark[1]&0&0&0&-       &1.4          &-       &-      &-          \\
  73\footnotemark[1] {\bf (C)}&0&5.0&0&-     &-            &$<1$    &$<1$   &-          \\
\end{tabular}
\end{ruledtabular}
\caption{Nominal compositions of all ingots used in this study, including in particular samples A (ingot \# 34, as-grown), B (ingot \# 37, annealed) and C (ingot \# 73, annealed) shown in Fig.~1 of the main manuscript. For each ingot, the table lists the alien phase content as estimated from powder X-ray diffraction measurements.}\label{XRD}
\footnotetext[1]{Samples grown with 4N yttrium}
\end{table*}

\begin{table*}[h]
\begin{ruledtabular}
\begin{tabular}{c*{6}{C}}
  Ingot No. \# &\multicolumn{3}{c}{EDS measured composition [at.\%]
                 }&\multicolumn{3}{c}{Atomic ratios
                                                                                      } \\
  \cmidrule(r){2-4} \cmidrule(r){5-7}
    (No. of spots probed)&Y&Fe&Ge&Fe:Ge&Fe:Y&Ge:Y\\
    \midrule
  27 (N=88)&20.01(5)&38.50(5)&41.50(3)&0.9276(15)&1.9240(51)&2.0741(51)\\
  57 (N=71)&19.95(5)&38.56(4)&41.50(4)&0.9292(13)&1.9330(56)&2.0803(60)\\
  34 (N=66)&19.92(5)&38.92(5)&41.17(4)&0.9453(15)&1.9537(54)&2.0667(56)\\
  73 (N=59)&19.85(5)&38.98(5)&41.17(6)&0.9469(18)&1.9635(65)&2.0736(70)\\
  \midrule
  FeGe$_2$ (N=27)&-&32.46(3)&67.54(3)&0.9611(10)$\times\frac{1}{2}$&-&-\\
  Y$_5$Ge$_3$ (N=39)&60.56(6)&-&39.44(6)&-&-&2.1710(40)$\times\frac{3}{10}$\\
\end{tabular}
\end{ruledtabular}
\caption{EDS data for four ingots and two reference compounds. Ingots \#27 and \#57 have produced annealed samples with very low RRR $\simeq 30$, whereas ingots \#34 and \#73 have produced annealed samples with RRR $> 150$. By normalising the EDS estimates for the Fe/Ge ratio against the EDS ratio measured in the case of the line compound FeGe$_2$, we can extract accurate estimates of the true Fe/Ge ratio in the four ingots, and these ratios are presented in 
  the main paper.}
\label{EDS}
\end{table*}

\end{document}